\documentclass[twocolumn,preprintnumbers,amsmath,amssymb,prl,superscriptaddress]{revtex4}
\usepackage{graphicx}
\usepackage{epstopdf}
\usepackage{dcolumn}
\usepackage{bm}
\usepackage{amsmath}
\usepackage{amssymb}
\usepackage{mathrsfs}
\usepackage{amsfonts}
\usepackage{times}
\usepackage{subfigure}
\usepackage{amstext}
\usepackage[colorlinks=true, citecolor=blue]{hyperref}
\newcommand{\beqa}{\begin{eqnarray}}
\newcommand{\eeqa}{\end{eqnarray}}

\newcommand{\ket}[1]{|#1\rangle}

\newcommand{\ann}[2][]{\hat #2^{#1}}
\newcommand{\crea}[2][]{\hat #2^{\dagger #1}}
\newcommand{\Osigma}[2]{\hat{\sigma}_{#1}^{#2}}

\begin{document}
\title{Universal spectral features of ultrastrongly coupled systems}
\author{Simone Felicetti}
\email{felicetti.simone@gmail.com}
\affiliation{Departamento de Física Teórica de la Materia Condensada and Condensed Matter
Physics Center (IFIMAC), Universidad Autónoma de Madrid, E- 28049 Madrid, Spain}
\affiliation{Laboratoire Mat\'eriaux et Ph\'enom\`enes Quantiques,
Universit\'e de Paris, CNRS, 75013 Paris, France}

\author{Alexandre Le Boit\'e}
\email{alexandre.leboite@univ-paris-diderot.fr}
\affiliation{Laboratoire Mat\'eriaux et Ph\'enom\`enes Quantiques,
Universit\'e de Paris, CNRS, 75013 Paris, France}

\begin{abstract}
We identify universal properties of the low-energy subspace of a wide class of quantum optical models in the ultrastrong coupling limit, where the coupling strength dominates over all other energy scales in the system.  We show that the symmetry of the light-matter interaction is at the origin of a two-fold degeneracy in the spectrum. We prove analytically this result for bounded Hamiltonians and extend it to a class of models with unbounded operators. As a consequence, we show that the emergence of superradiant phases previously investigated in the context of critical phenomena, is a general property of the ultrastrong coupling limit. The set of models we consider encompasses different scenarios of possible interplay between critical behaviour and superradiance.
\end{abstract}

\maketitle
The experimental control of the coherent interaction between light and matter is one of the corner stones of the recent developments in the field of quantum technologies. Experiments in cavity quantum electrodynamics (cavity QED) have been essential both for our understanding of quantum-optical phenomena at the most fundamental level ~\cite{Haroche:2006, Gleyzes:2007} and for the implementation of quantum information protocols~\cite{Imamoglu:1999,Majer:2007}. A decisive challenge in cavity QED experiments consists in increasing the strength of the coupling between light and matter. In this respect, two main milestones have been reached, each of them leading to new features and potentially new technological functionalities~\cite{Frisk-Kockum:2019}. A key step was the achievement of the  strong coupling regime, where the coupling strength is larger than any dissipation rate.  This regime has been demonstrated in atomic cavity QED \cite{Rempe:1987}, semiconductor nanostructures \cite{Reithmaier:2004, Peter:2005} and superconducting circuits \cite{Wallraff:2004}, leading to the observation of genuine quantum effects such as vacuum Rabi oscillations and photon antibunching~\cite{Imamoglu:1997, Birnbaum:2005, Bozyigit:2010, Lang:2011, Hoffman:2011}. 

In the last decade, we entered in a new era of cavity QED with the achievement~\cite{Todorov:2010,Niemczyk:2010,Forn-Diaz:2010,Forn-Diaz:2017,Yoshihara:2017,Chen:2017}  of the ultrastrong coupling (USC) regime, where the coupling strength becomes comparable or even larger than the cavity frequency  \cite{Devoret:2007, Bourassa:2009,   Forn-Diaz:2016}.  Furthermore, recently developed quantum simulation techniques made it possible to observe~\cite{Langford:2017, Braumuller:2017,Markovic:2018, Peterson:2019} the physics of the ultrastrong coupling regime even in systems that do not naturally achieve the required interaction strength. The rich phenomenology of this new regime of cavity QED has been the focus of an intense research activity. The USC regime proved to induce profound modifications in a variety of fundamental quantum optical phenomena, ranging from vacuum radiation~\cite{DeLiberato:2009, Sanchez-Burillo:2019} to single-photon emission~\cite{Ridolfo:2012, LeBoite:2016}, scattering processes~\cite{Shi:2018} and transport properties~\cite{Felicetti:2014, Bartolo:2018}.
Among its prominent features, it was also recognized that some systems exhibit a two-fold degenerate ground state in the USC regime~\cite{Nataf:2010, Casanova:2010, Rossatto:2017}. It was proposed to exploit this interesting feature for the design of protected qubits~\cite{Ashhab:2010, Nataf:2011, Kyaw:2015}.

Ultrastrong light-matter interactions in cavity QED may also give rise to superradiant phase transitions (SPT)~\cite{Larson:2017, Peng:2019}. From a theoretical point of view, the Dicke model is a paradigmatic example in which such a phase transition occurs in the thermodynamical limit, when the number of atoms coupled to the cavity mode is going to infinity~\cite{Kirton:2019}. More recently, a SPT have been predicted to occur also in  the quantum Rabi model~\cite{Hwang:2015}, which is a finite-component model. In finite-component models the thermodynamical limit can be defined formally by letting one parameter of the Hamiltonian go to infinity. In addition to a macroscopic number of photons in the ground state,  the superradiant phase is in both cases characterized by a two-fold degeneracy of the low-energy eigenstates and a breaking of the parity symmetry.  Note that violation of gauge invariance ~\cite{DeBernardis:2018, Stokes:2019, DiStefano:2019} and the role of the usually-neglected diamagnetic ``$A^2$-term''~\cite{Nataf:2010b, Garcia-Ripoll:2015, Manucharyan:2017, DeBernardis:2018b} can constrain the validity of effective models in the USC regime. Nevertheless, Hamiltonian engineering via parametric couplings or analog quantum simulation schemes makes it possible to observe superradiant phase transitions and to feasibly explore extreme regimes of parameters.

In this letter, we show that two-fold degeneracy and parity-symmetry breaking are universal properties of quantum optical models in the ultrastrong coupling limit, where the coupling strength dominates over all other energy scales.
We give a general proof of this result in the case of bounded Hamiltonians and extend it to a set of models with unbounded operators. The class of Hamiltonians we consider includes coupled non-linear oscillators, such as Bose-Hubbard chains, which are relevant for a wide class of experimental platforms. We show that in such bosonic systems a superradiant phase always emerges in the ultrastrong coupling limit, whether in the form of a crossover or a phase transition. In particular, the phenomenology of the SPT occurring in both the Rabi and Dicke models is recovered by introducing proper scalings of the parameters. Finally, we show that a novel interplay between critical behavior and supperradiance can emerge in the ultrastrong coupling limit.

\begin{figure}[]
\begin{center}
\includegraphics[width=0.45\textwidth]{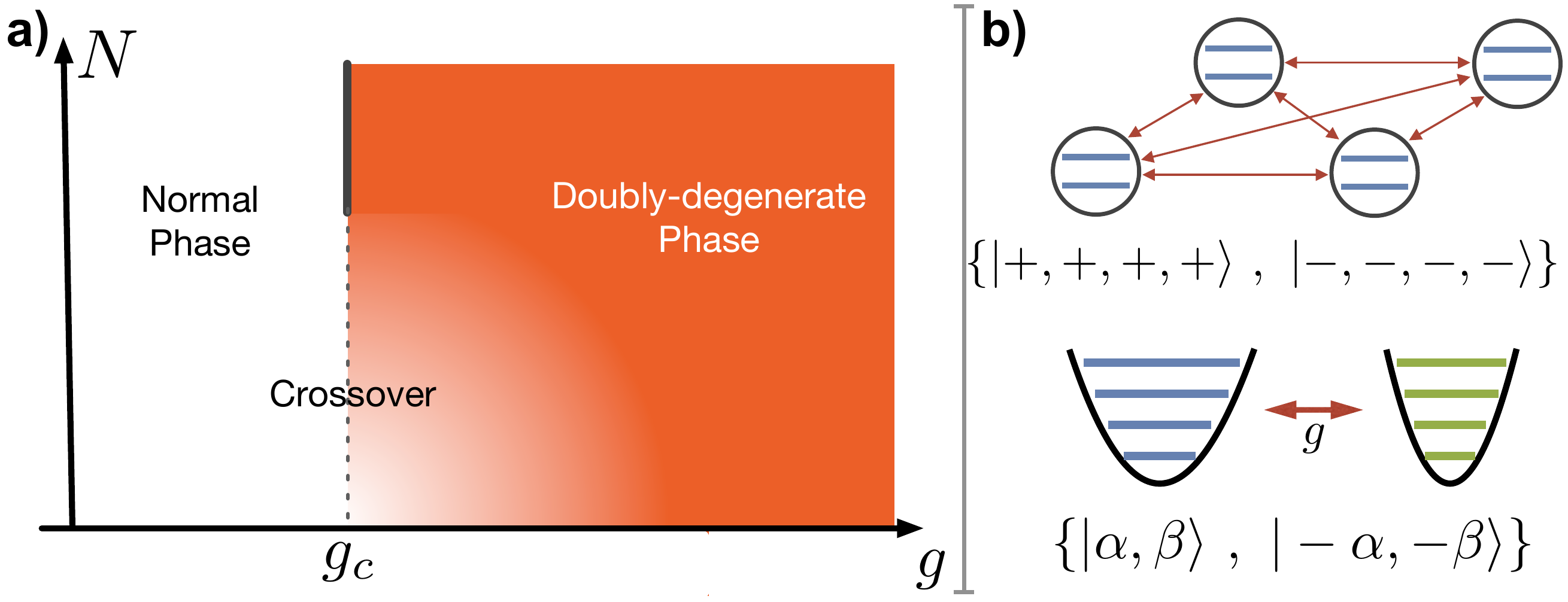}
\caption{a) General structure of the phase diagram for the considered class of quantum light-matter interaction models. The variable $N$ is an arbitrary scaling parameter defining an effective thermodynamical limit. As the coupling strength is increased, the system can enter a crossover region or undergo a critical phase transition. A doubly-degenerate parity-breaking phase always emerges in the ultrastrong coupling limit ($g \to \infty $ ).  b) Example of the doubly-degenerate subspace for a  finite-dimensional system and for a Bose-Hubbard dimer. The states $|\pm\rangle$ are eigenstates of the coupling operators of Eq.~(\ref{multiBounded}) and the displacement parameters $\alpha$ and $\beta$ are solutions of Eq. (\ref{NoH1}).} \label{fig:1}
\end{center}
\end{figure}

\paragraph{Bounded operators}
When the Hamiltonian is bounded, the proof of the result mentioned above is straightforward but the intuition it provides is nonetheless useful. Consider the Hilbert space $\mathcal{H} = \mathcal{H}_1\otimes\mathcal{H}_2$ of two coupled parity-conserving systems, with the following total Hamiltonian
\begin{equation}
H = H_1 + H_2 + g H_I.
\end{equation}
Local parity conservation is expressed as $[\mathcal{P}_1, H_1] = [\mathcal{P}_2,H_2] = 0$.
A key point is the symmetry properties of the interaction term: $H_I = X_1 \otimes X_2$. The operators $X_1$ and $X_2$ are generic Hermitian operators acting on $\mathcal{H}_1$ and $\mathcal{H}_2$ respectively,  but we assume that  they satisfy the following anticommutation relations 
\begin{equation}\label{AntiCom}
\{\mathcal{P}_1,X_1\} = \{\mathcal{P}_2,X_2\} = 0,
\end{equation}
which is valid, e.g. for all quantum optical models involving dipolar light-matter coupling. As a result, the Hamiltonian commutes with a global parity symmetry operator, $[H,\mathcal{P}_1\mathcal{P}_2] = 0.$  For now we also assume that there is no dark state in the system, i. e. $\mathrm{Ker}(X_i) = \emptyset $.
In all that follows, we define the ultrastrong coupling limit as $H_{USC} = \lim_{g\to \infty} H/g$. This condition can be viewed as a formal thermodynamical limit for finite-component systems. When all the operators are bounded, this limit is well defined and the total Hamiltonian can be approximated as
\begin{equation}
\label{H_USC}
H/g \approx H_{USC} = X_1\otimes X_2. 
\end{equation}
The eigenstates of $H_{USC}$ are product states of the form $|\Psi\rangle = |\phi_1\rangle|\phi_2\rangle$,
where
$X_i|\phi_i\rangle = E_i|\phi_i\rangle$.
From Eq.~\eqref{AntiCom} and Eq.~\eqref{H_USC} we see that the states $|\Psi\rangle$ and $\mathcal{P}_1\mathcal{P}_2|\Psi\rangle$ are degenerate and orthogonal. In this degenerate subspace, the superpositions $|\Psi_+\rangle =  \frac{1}{\sqrt{2}}(|\Psi\rangle + \mathcal{P}_1\mathcal{P}_2|\Psi\rangle)$, $|\Psi_-\rangle =  \frac{1}{\sqrt{2}}(|\Psi\rangle - \mathcal{P}_1\mathcal{P}_2|\Psi\rangle)$ are eigenstates of the total parity operator $\mathcal{P}_1\mathcal{P}_2$. This proves that in the USC limit, the spectrum is at least two-fold degenerate and that the degenerate subspaces contain eigenstates with opposite parity, as illustrated in Fig.~\ref{fig:1}. We refer to the latter feature as symmetry breaking. The generalization to a multipartite system can be obtained in a similar fashion: consider now a general Hilbert space $\mathcal{H} = \bigotimes_i \mathcal{H}_i$ with
\begin{equation}\label{multiBounded}
H = \sum_i H_i + \sum_{i>j}g_{ij}X_i\otimes X_j.
\end{equation}
where the symmetry assumption takes the form
$\forall i, [\mathcal{P}_i, H_i] = 0$ and $\forall i, \{\mathcal{P}_i, X_i\} = 0$. The same arguments used in the bipartite case lead to the conclusion that the eigenstates are of the form
$|\Psi_+\rangle = \frac{1}{\sqrt{2}}\left(\bigotimes_i |\phi_i\rangle + \bigotimes_i \mathcal{P}_i|\phi_i\rangle \right)$ and
$|\Psi_-\rangle = \frac{1}{\sqrt{2}}\left(\bigotimes_i |\phi_i\rangle - \bigotimes_i \mathcal{P}_i|\phi_i\rangle \right)$,
which are degenerate and are also eigenstates of the total parity operator $\mathcal{P} = \prod_i \mathcal{P}_i$, with opposite parity. The structure of the low-energy spectrum is shown in Fig.~\ref{fig:1}.

\paragraph{Unbounded case}
When the Hamiltonian is unbounded, the operator $H_{USC}$ defined in Eq.~(\ref{H_USC}) may lead to unphysical results. As an example, consider two coupled harmonic oscillators and the following Hamiltonian
\begin{equation}
H_0 = \omega_1 \crea{a}\ann{a} + \omega_2 \crea{b}\ann{b} + g(\crea{a} + \ann{a})(\crea{b} + \ann{b})
\end{equation}
The parity of the number of excitations is conserved and the coupling operators satisfy the requirements of Eq. (\ref{AntiCom}). However, $H_I = g(\crea{a} + \ann{a})(\crea{b} + \ann{b})$ is not lower-bounded, leading to a dynamical instability of the system for $g>\frac{\sqrt{\omega_1\omega_2}}{2}$. A possible way to stabilize such a system in the USC limit is to add a nonlinear quartic term to each oscillator. Hence, we begin by extending the previous results to Bose-Hubbard dimers, described by the following Hamiltonian, 
\begin{equation}
H = H_0  + \epsilon_1 \crea{a}\crea{a}\ann{a}\ann{a} +\epsilon_2\crea{b}\crea{b}\ann{b}\ann{b},
\label{boseHdimers}
\end{equation}
where $\epsilon_1$ and $\epsilon_2$ may take arbitrary values. Note that in the context of quantum optics, such models have been widely used to described various nonlinear elements, and they are relevant to many experimental platforms~\cite{Carusotto:2013}.  
In the USC limit, a low-energy effective Hamiltonian can be found by applying a displacement operator on both fields $\hat{a}$ and $\hat{b}$. The displaced Hamiltonian is expressed as $H_{\alpha,\beta} = D_a^\dagger[\alpha]D_b^\dagger[\beta] H D_a[\alpha]D_b[\beta]$,
where $D_c[\gamma] = e^{\gamma\crea{c} - \gamma^*\ann{c}}$ with $c \in \{a,b\}$ and $\gamma \in \mathbb{C}$. 
This operation is meant to displace  the vacuum state into a new local minimum in the effective potential landscape. Accordingly, we choose the values of the displacements $\alpha$ and $\beta$ for which the linear terms appearing in $H_{\alpha, \beta }$ cancel out.
This condition is set by the system of equations
\begin{equation}\label{NoH1}
\begin{cases}
\omega_1\alpha+2\epsilon_1\alpha|\alpha|^2 + g(\beta^*+\beta) = 0\\
\omega_2\beta+2\epsilon_2\beta|\beta|^2 + g(\alpha^*+\alpha) = 0.
\end{cases}
\end{equation}
Apart from linear terms, the Hamiltonian $H_{\alpha,\beta}$ in its general form contains the following additional parts:
(i) harmonic terms proportional to $ \epsilon_1|\alpha|^2$, $\epsilon_2|\beta|^2$ that renormalize the oscillators frequency, (ii) squeezing terms proportional to $ \epsilon_1\alpha^2$, $\epsilon_2\beta^2$,
(iii) additional third order terms proportional to $\epsilon_1\alpha, \epsilon_2\beta$.
Before looking at $\lim_{g\to\infty} H_{\alpha,\beta}/g$ and find $H_{USC}$, let us mention a general feature of $H_{\alpha,\beta}$. Provided that a meaningful solution to Eq.~(\ref{NoH1}) is found, the quadratic part of the displaced Hamiltonian can always be cast into the following form 
\begin{align}\label{Hdis2}
H^{(2)}_{\alpha,\beta} = 2g|x|\crea{a}\ann{a} + \frac{2g}{|x|}\crea{b}\ann{b} +g(\crea{a} + \ann{a})(\crea{b}+ \ann{b}) +  \nonumber \\
 (-\frac{\omega_1}{2} + g|x|)(\crea{a} + \ann{a})^2 + (-\frac{\omega_2}{2} + \frac{g}{|x|})(\crea{b} + \ann{b})^2, 
\end{align}
where $x = \beta/\alpha$.
Without the squeezing terms, this Hamiltonian would always lead to a dynamical instability, but the additional  squeezing terms, which are positive by construction, stabilize it. The quadratic part of the Hamiltonian displaced according to Eq.~\eqref{NoH1} is therefore always well defined.
In the USC limit, we find an approximate solution to Eq.~\eqref{NoH1} where $\alpha^2$ and $\beta^2$ are proportional to $g$, that is $\alpha^2 = \frac{g}{\sqrt[4]{\epsilon_1^3\epsilon_2}}$ and $\beta = -\alpha\left(\frac{\epsilon_1}{\epsilon_2}\right)^{1/4} $.
As a result, the limit $g\to \infty$ is well defined and all non-quadratic terms of $H_{\alpha,\beta}$ can be neglected. The resulting Hamiltonian depends only on a single parameter $\eta =\left(\frac{\epsilon_1}{\epsilon_2}\right)^{1/4}$,
\begin{align}
H^{(\alpha,\beta)}_\mathrm{USC} = 2\eta \crea{a}\ann{a} + \frac{2}{\eta}\crea{b}\ann{b} +\eta(\ann{a} + \crea{a})^2 + \frac{1}{\eta}(\ann{b} + \crea{b})^2 +  \nonumber \\+ (\crea{a} + \ann{a})(\crea{b} + \ann{b}).
\end{align}
From this last expression we conclude that structure of the spectrum identified above in the case of bounded operators, also extends to the low-energy sector of Bose-Hubbard dimers, as shown in Fig.~\ref{fig:1}. Here, the degeneracy in the spectrum comes from Eq. (\ref{NoH1}), which is invariant under the transformation $\{ \alpha, \beta \} \to \{-\alpha, -\beta\}$. In addition, the ground state is superradiant, in the sense that the parameters $\alpha, \beta$ are both proportional to $g$ and both modes are therefore macroscopically occupied in the USC limit.

\begin{figure}[]
\begin{center}
\includegraphics[width=0.45\textwidth]{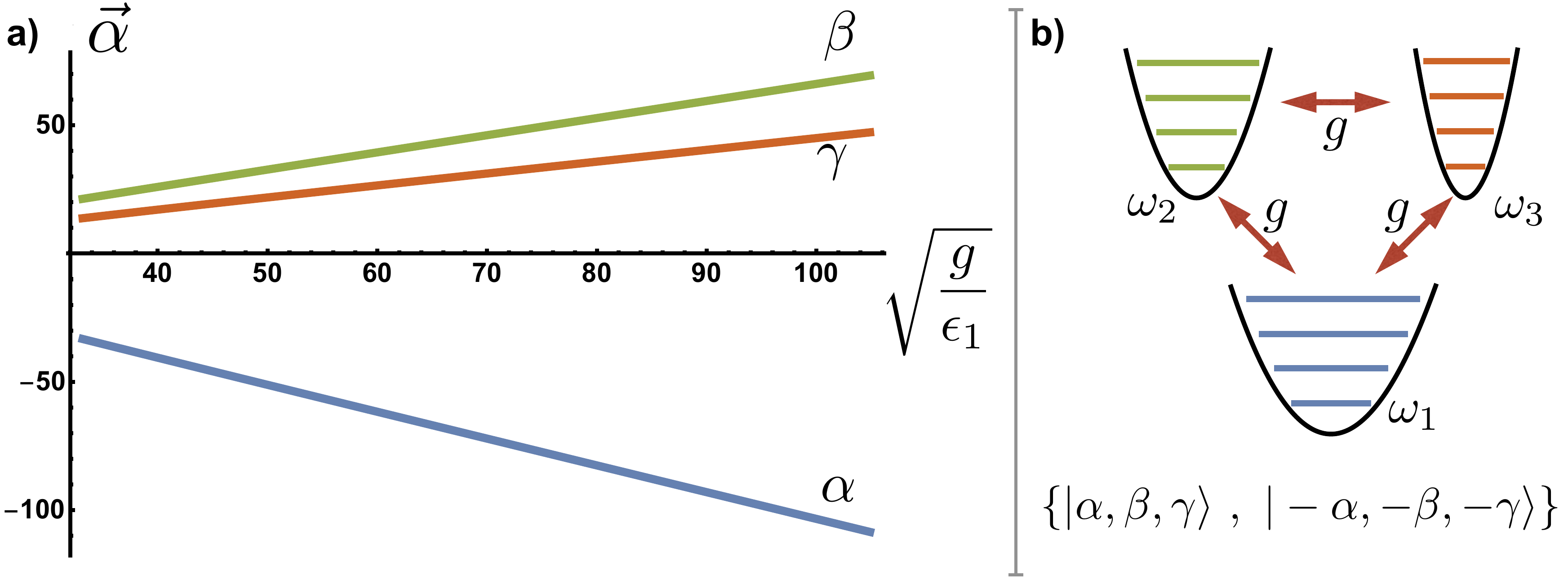}
\caption{\label{fig:2} a) Values of the displacements  $\vec{\alpha}$, solution of Eq.~\eqref{NoH1_tri}, for increasing values of the homogeneous coupling strength $g_i = g$, for $\omega_i=\{1,2,4 \}$ ,  $\epsilon_i = \omega_i\times 10^{-2}$. b) The ground state of a Bose-Hubbard trimer in the USC limit is described by quantum fluctuations on top of macroscopic displacements in each oscillator.
}
\end{center}
\end{figure}

As in the bounded case, let us show that we can generalize the results to  multipartite systems with a continuum spectrum.
Let us consider the Bose-Hubbard chain
\begin{equation}
\label{BHchain}
H = \sum_{i} \left( \omega_i \crea{a}_i\ann{a}_i + \epsilon_i \crea{a}_i\crea{a}_i\ann{a}_i\ann{a}_i \right) + H_I, 
\end{equation}
where the interaction Hamiltonian is given by  linear two-body coupling terms
$H_I = \sum_{i>j} g_{i,j}(\crea{a}_i + \ann{a}_i)(\crea{a}_j + \ann{a}_j)$.
To find an effective low-energy description in the USC limit, we apply the unitary transformation $H_{\vec{\alpha}} = D^\dagger[\vec{\alpha}]  H^{\rm t} D[\vec{\alpha}] $, where we defined $D[\vec{\alpha}] = \bigotimes_i D_i[\alpha_i]$, being $D_i^\dagger[\cdot]$  a displacement operator applied on the field $\ann{a}_i$.
As in the bipartite unbounded case, we look for  displacement parameters $\vec{\alpha}$ for which the local linear terms vanish. The generalization of Eq.~\eqref{NoH1} to the multipartite case corresponds to
\begin{equation}
\label{NoH1_tri}
\omega_i \alpha_i +2\epsilon_a \alpha_i^3 + \sum_{j\neq i} 2 g_{ij}\alpha_j = 0, \quad \forall\ i.
\end{equation}
For each vector of displacement parameters $\vec{\alpha}$ that satisfies this equation, we obtain an effective Hamiltonian of the form,
\begin{equation}
\label{Halphabetagamma}
H_{\vec{\alpha}} = 
\sum_i \left[ \Omega_i \crea{a_i}\ann{a_i} 
+ \epsilon_i \alpha_i^2  \left(\hat{a}_i^{\dagger 2} +\hat{a}_i^2 \right)  \right] + H_I^{\rm t} + E_0,
\end{equation}
where we defined $\Omega_i =  \omega_i + 4\epsilon_i \alpha_i^2$. 
The constant energy term $E_0 = \sum_i \omega_i|\alpha_i|^2 + \epsilon_i |\alpha_i|^4 + 4\sum_{i>j} g \alpha_i \alpha_j$ depends on the displacement parameters and must be kept into account to determine the eigenenergies of the effective Hamiltonian.
We will first describe  analytical results for a Bose-Hubbard chain of three cavities in the fully-symmetric case $\omega_i = \omega$, $\epsilon_i = \epsilon$ and $g_{i,j} = g$, for any $\{i,j\}$. We look for displacements parameters $\vec{\alpha} = \{\alpha,\beta,\gamma\}$ that satisfy Eq.~\eqref{NoH1_tri}. 
Approximated solutions in the USC limit can be found  taking the ansatz $\alpha,\beta,\gamma \sim \sqrt{g}$, which implies that the three fields are simultaneously displaced and acquire a number of excitations proportional to the coupling strenght.  The value of the displacements can be expressed as $\left\{ \alpha_\pm, \beta_\pm, \gamma_\pm \right\} = \left\{\pm k \sqrt{\frac{g}{\epsilon}},\pm k \sqrt{\frac{g}{\epsilon}}, \mp k(k+k^3)\sqrt{\frac{g}{\epsilon}} \right\}$, where the parameter $k\approx 0.74$ is real and the solution of the equation $k^{(8/3)} + k^{(2/3)} - \sqrt[3]2 = 0$. 
Notice that in the USC limit the stability of the displaced Hamiltonian~\eqref{Halphabetagamma} is granted by the renormalization of the frequencies $\Omega_i$, and so only solutions of Eq.~\eqref{NoH1_tri} in which all fields are displaced give rise to stable Hamiltonians. The two displacement vectors $\pm \vec{\alpha}$ lead to the same effective Hamiltonian, and so we proved that the low-energy eigenspectrum of a symmetric three-site Bose-Hubbard chain for $g\rightarrow \infty$ is  composed of degenerate doublets with opposite parity. As illustrated in Fig.\ref{fig:2}(b) the eigenvectors are given by displaced Fock states $D_a[\alpha_\pm]D_b[\beta_\pm]D_c[\gamma_\pm] \ket{n_1,n_2,n_3}$.
  Notice that in the fully-symmetric case any permutation of  $\{\alpha, \beta, \gamma \}$  provides an equivalent set of solutions. This additional degeneracy is due to permutation symmetry and it is expected to be lifted as soon as inhomogeneities are introduced. 
To verify this intuition we take a semi-analytic approach to determine the low-energy eigenspectrum in the non-symmetric case. First, we numerically solve Eq.~\eqref{NoH1_tri} to determine the values of the displacement parameters that lead to a quadratic Hamiltonian of the form of Eq.~\eqref{Halphabetagamma}, and then we derive the coefficients of the Bogoliubov transformation that diagonalizes the resulting Hamiltonian. The results of this analysis (see Fig.~\ref{fig:2}) show that the eigenspectrum in the inhomogeneous case is given by degenerate doublets with opposite parity, while the permutation symmetry is indeed lifted. In the system ground state, each bosonic mode is displaced by an amount proportional to $\sqrt{g/\epsilon}$, as in the symmetric case. 
\paragraph{Finite-component phase transitions}~  We have seen that the ground state of the effective bosonic Hamiltonians derived above are superradiant in the USC limit. However, there is no phase transition in this case since the ground state properties of $H_{\mathrm{USC}}$ vary smoothly with $g$ (see Fig.~\ref{fig:1}). In the following we show that by defining another type of thermodynamic limit, i.e. a specific scaling of the parameters with a formal macroscopic factor $N$, the model can reproduce different kinds of superradiant phase transitions, including both types associated with the Dicke and quantum Rabi models. More precisely, considering the Bose-Hubbard dimers of Eq.~\eqref{boseHdimers}, we define Dicke-type superradiance~\cite{Kirton:2019} as a phase in which both oscillators acquire a macroscopic coherence in the superradiant phase. In contrast, we call Rabi-type superradiance~\cite{Hwang:2015} a superradiant phase in which only one of the oscillator gains macroscopic coherence. To capture  a Dicke-type superradiant phase, we are therefore looking for solutions of Eq.~(\ref{NoH1}) in which the parameters $\alpha$ and $\beta$ are both scaling as $\sqrt{N}$. A relevant scaling to find this type of solution is the following
\begin{align}
H_d(N,\lambda) &= \omega_1 \crea{a}\ann{a} + \omega_2 \crea{b}\ann{b}+g(\crea{a} + \ann{a})(\crea{b} + \ann{b})\nonumber \\
&+ \frac{\epsilon_1}{N} \crea{a}\crea{a}\ann{a}\ann{a} +\frac{\epsilon_2}{N}\crea{b}\crea{b}\ann{b}\ann{b},
\end{align}
where we consider the limit $N \to \infty$ for different values of the parameters. Here, the parameter driving the phase transition is $\lambda = \frac{2g}{\sqrt{\omega_1\omega_2}}$.
For $\lambda < 1$, the system is in the ``normal phase''. In the limit $\lim_{N \to \infty} H_d(N,\lambda)$, the Hamiltonian obtained simply by neglecting the terms proportional to $1/N$ is quadratic and well defined. The ground state is a squeezed vacuum obtained after a standard Bogoliubov transformation.
For $\lambda > 1$, the phase is superradiant: there is a solution to Eq. (\ref{NoH1}) with $\alpha \sim \beta \sim \sqrt{N}$. In the displaced Hamiltonian, since $H^{(3)}_{\alpha,\beta} \sim \frac{\epsilon_1}{N}\alpha$ and $H^{(4)}_{\alpha,\beta} \sim \frac{\epsilon_1}{N}$, all non quadratic terms vanish for $N \to \infty$. The Hamiltonian is then given by Eq (\ref{Hdis2}).
For a Rabi-type superradiance, the solutions we are interested in scale as $\alpha \sim \sqrt{N}$  and $ \frac{\beta}{\alpha} \sim \frac{1}{\sqrt{N}}$. A scaling providing a superradiant phase with this properties is given by
\begin{align}
H_r(N,\lambda) &= \omega_1 \crea{a}\ann{a} + N\omega_2 \crea{b}\ann{b}+\sqrt{N}g(\crea{a} + \ann{a})(\crea{b} + \ann{b}) \nonumber \\
&+ \frac{\epsilon_1}{N} \crea{a}\crea{a}\ann{a}\ann{a} +\frac{\epsilon_2}{N}\crea{b}\crea{b}\ann{b}\ann{b}.
\end{align}
Once again, the scaling is such that $\lambda$ is finite and governs the transition. The normal phase can be treated in the same way as for the Dicke-type phase considered above. Given the infinite detuning between the two oscillators in the limit $N \to \infty$, exact diagonalization~\cite{Emary:2003} shows that the ground state is trivial in this case (decoupled vacuum state). Howerver, when $\lambda > 1$ we have, for $N \to \infty$,
\begin{align}
\frac{\beta}{\alpha} =& -\frac{1}{\sqrt{N}}\frac{2g}{\omega_2},\\
\alpha^2 =& N \frac{\omega_1(\lambda^2 -1)}{2\epsilon_1}.
\end{align}
This defines a superradiant phase, whose effective Hamiltonian is
\begin{align}
H_{eff} &= \omega_1\lambda^2\crea{a}\ann{a} + \frac{\omega_1}{2}(\lambda^2-1)(\crea{a} + \ann{a})^2 +  N\omega_2 \crea{b}\ann{b} \nonumber \\ 
&+ \sqrt{N}g(\crea{a} + \ann{a})(\crea{b} + \ann{b}).
\end{align}
As previously, the higher-order terms of the effective Hamiltonian vanish and the two oscillators are decoupled in the limit $N \to \infty$. 
This result shows that the Rabi-type superradiant phase transition is not specific to the quantum Rabi model but can also occur in a finite-component system with vanishingly-small nonlinearities.

Let us now show that in the USC limit a first-order phase transition can emerge between two phases that are both two-fold degenerate with broken parity symmetry.
We consider an extended version of a Dicke model where, in addition to standard light-matter coupling, we include an interatomic interaction. We restrict ourselves to the two-atom case, but the properties we discuss can be directly generalized to any finite number of atoms. The system Hamiltonian is given by $H^{2D} =  \omega \crea{a} \ann{a} + \frac{\omega_q}{2} \left( \Osigma{z}{1}+ \Osigma{z}{2} \right) + H_I^{2D}$, where  $H^{2D}_I = g\sqrt{N} \left(  \crea{a} + \ann{a}  \right)\left(  \Osigma{x}{1} + \Osigma{x}{2}  \right) + \chi N\Osigma{x}{1}  \Osigma{x}{2}$, where we have already introduced an effective scaling parameter $N$.
In the USC limit, here obtained for $N\rightarrow \infty$, the local atomic energy terms can be neglected, as they correspond to bounded operators with a finite prefactor. As a result, in this limit the Hamiltonian $H^{2D}$ is block-diagonal  and can be rewritten as, 
\begin{equation}
H^{2D}_{s_1,s_2} =  \omega \crea{a} \ann{a} + g\sqrt{N}\left(s_1 + s_2  \right)\left(  \crea{a} + \ann{a}  \right) - \chi N s_1 s_2,
\end{equation}
where $s_1$ and $s_2$ are the corresponding eigenvalues of  $\Osigma{x}{(1)}$ and  $\Osigma{x}{(2)}$, respectively. Each block of the Hamiltonian is diagonalized by the displacement $D_a[\alpha]$ where $\alpha = -\frac{g\sqrt{N}}{\omega}\left( s_1 + s_2 \right)$.
The low-energy eigenspectrum in the USC limit is  then composed of degenerate doublets of two different kinds, obtained for parallel ($s_1 = s_2$) or orthogonal ($s_1 = -s_2$) atomic spins. For parallel spins, we obtain an effective low-energy Hamiltonian $H_\parallel = \omega \crea{a} \ann{a} + \frac{4g^2 N}{\omega} -\chi N $ and the ground-state energy $E_0^\parallel = \frac{4g^2 N}{\omega} -\chi N$.
On the other hand, for orthogonal spins we obtain $H_\perp = \omega \crea{a} \ann{a} + \chi N $, with two-fold degenerate ground state $\ket{g_\perp} \in \left\{\ket{+,-,0},\ket{-,+,0}  \right\}$ and
ground-state energy $E_0^\perp = \chi N$.
By looking at the ground-state energies we see  that the system undergoes a first-order quantum phase transition for $\chi_c = \frac{4g^2}{\omega}$. For small  values of $\chi$, the ground state is in the orthogonal-spin subspace and it is given by  $\ket{g_\perp} \in \rm{span}\left\{\ket{+,-,0},\ket{-,+,0}  \right\}$. As $\chi$ is increased above the critical value, the groundstate switches to the parallel-spin sector and it is given by a superradiant state  $\ket{g_\parallel} \in \rm{span}\left\{\ket{+,+,-\alpha},\ket{-,-,\alpha}  \right\}$, in which the bosonic field acquires a number of photons proportional to $\alpha^2 = 4N(g/\omega)^2$. This quantum phase transition is due to the competition between the inter-atomic coupling and the light-matter interaction. As the latter pushes the atoms towards an antiferromagnetic configuration, the phase transition occurs for ferromagnetic inter-atomic interactions $\chi>0$. Notice that this kind of phase transition has been identified so far only for a similar system in the thermodynamical limit~\cite{DeBernardis:2018b}.

To conclude, we have shown that in the ultrastrong-coupling limit the low-energy spectra of quantum-optical models are characterized by a universal structure composed of degenerate doublets that break the parity symmetry.
Furthermore, we have illustrated how in finite-component systems the features of the USC limit can emerge either as the result of a smooth crossover or as critical phase transitions. Finally, we have shown that in the USC limit a first-order quantum phase transition can take place between two different kinds of parity-broken phases.
The framework we introduced paves the way to the thorough phenomenological study of the USC limit of different light-matter interaction models.
Its extension to open systems, in which finite-component dissipative phase transitions have been predicted~\cite{Bartolo:2016, Casteels:2017, Hwang:2018}, offers interesting perspectives.

\paragraph{} -- We thank Alberto Biella and Cristiano Ciuti for useful discussions. S. F. acknowledges support from the European Research Council (ERC-2016-STG-714870).

\end{document}